\documentclass[11pt]{amsart}
\usepackage{fullpage}
\usepackage{amsthm,amsmath,amssymb}
\usepackage[foot]{amsaddr}

\usepackage{enumitem}
\usepackage{booktabs}
\usepackage{graphicx}
\usepackage{subcaption}
\usepackage{tikz}
\usepackage{natbib}

\usepackage[colorlinks=true,linkcolor=red,breaklinks=true,citecolor=blue,urlcolor=cyan]{hyperref}

\graphicspath{{./figs/}} 

\theoremstyle{remark}

\title{Temporal discontinuity trials and randomization: success rates versus design strength}

\author{Brian Knaeble}
\address{Department of Computer Science, Utah Valley University, Orem, UT}
\email{bknaeble@uvu.edu}

\author{Erich Kummerfeld}
\address{Institute for Health Informatics, University of Minnesota, Minneapolis, MN}
\email{erichk@umn.edu}

\keywords{Randomization, Experimental Control, Opposites, Discontinuity Design}

\begin{document}
\maketitle

\begin{abstract}
We consider the following comparative effectiveness scenario. There are two treatments for a particular medical condition: a randomized experiment has demonstrated mediocre effectiveness for the first treatment, while a non-randomized study of the second treatment reports a much higher success rate. On what grounds might one justifiably prefer the second treatment over the first treatment, given only the information from those two studies, including design details? This situation occurs in reality and warrants study. We consider a particular example involving studies of treatments for Crohn's disease. In order to help resolve these cases of asymmetric evidence, we make three contributions and apply them to our example. First, we demonstrate the potential to improve success rates above those found in a randomized trial, given heterogeneous effects. Second, we prove that deliberate treatment assignment can be more efficient than randomization when study results are to be transported to formulate an intervention policy on a wider population. Third, we provide formal conditions under which a temporal-discontinuity design approximates a randomized trial, and we introduce a novel design parameter to inform researchers about the strength of that approximation. Overall, our results indicate that while randomization certainly provides special advantages, other study designs such as temporal-discontinuity designs also have distinct advantages, and can produce valuable evidence that informs treatment decisions and intervention policy.
\end{abstract}

\section{Introduction}
A recent study by \citet{Chiba2017} reported a 96\% remission rate for patients with Crohn's disease. This is a staggering figure considering the 30-60\% remission rates reported for other treatments \citep{Colombel2010,Jongsma2020,Sands2022}. If externally valid and transportable to the global population, this treatment could benefit millions of people \citep{Alatab2020}. The validity of this study is questionable, however, as it did not use a randomized design. Its results may not be transportable to another population. If that is the case, studies attempting to replicate their success would at best be a waste of time and resources, and at worst would harm their patient populations with a treatment inferior to standard care.

In this paper we consider the study design employed by \citet{Chiba2017} and evaluate the conditions under which findings from studies like theirs might be transportable to other populations. We describe their protocol as a \emph{temporal-discontinuity} design, and investigate the theoretical properties of temporal-discontinuity studies. Throughout, we make comparisons to what is considered the benchmark standard for human studies, the randomized controlled trial (RCT).

For many researchers, randomization is the ``reasoned basis'' for causal inference \citep{Fisher}, and the context of that expression is randomization inference \citep[p. 37]{Rose10}. Randomized controlled trials are widely considered the gold standard of ``empirical biomedical investigation'', but that status is contested in this era of big data \citep{Jones2015}. A principle of evidence based medicine is that a hierarchy of evidence exists, but it is a changing hierarchy \citep{Murad2016}. According to \citet{Gerstman2023}, ``the gold standard design is whatever method provides the information you need in the most reliable way.'' 

The study by \citet{Chiba2017} is one such example of a study that may have provided the needed information in the most reliable way, however as we will show, the situation is more complicated. While it is an example of a highly successful non-randomized trial, their non-randomized study design has both benefits and drawbacks relative to a randomized design. The relative benefits of using a temporal-discontinuity design include the possibility of having direct transportability to other populations and having increased power relative to the size of the recruited population. The relative drawbacks include requiring an additional condition to be assumed or supported with data. If this condition is met, however, the results of the discontinuity-design study are as valid as the results of a randomized design, and it is plausible that this condition could be met in some studies, including the study by \citet{Chiba2017}.

The remainder of this paper is organized as follows. In Section \ref{methods} we introduce our notation for analyzing success rates and transportability. We utilize that notation to define randomized and temporal-discontinuity trials, and we formulate a constrained optimization problem for sensitivity analysis. In Section \ref{results} we describe three results to support analysis of the comparative effectiveness of various treatments. The first result demonstrates the primacy of experimental control over randomization. It shows how deliberate treatment assignment is more efficient for transport than randomized treatment assignment. The second result introduces a formal condition for equivalence between temporal-discontinuity and randomized designs. The third result is a solution to the constrained optimization problem, and it can be used to assess the sensitivity of success rates to departures from the assumption of homogeneous effects. In Section \ref{application} we apply our results to analyze the comparative effectiveness of three popular dietary interventions for Crohn's disease, including the intervention studied by \citet{Chiba2017}. Further discussion is presented in Section \ref{discussion}. Mathematical proofs and supplementary material are provided in the appendix. 
\subsection{Our contributions}
Our novel contributions in this paper are the following.
\begin{enumerate} 
    \item We demonstrate potential for conditional interventions to increase the success rate above the maximum success rate of treatments studied in a randomized experiment, thus demonstrating non optimality of formulating intervention policy based solely on a randomized trial.
    \item We demonstrates how deliberate treatment assignment can be more efficient than randomization when study results are to be transported to formulate an intervention policy on a wider population.
    \item We provide formal conditions under which the temporal-discontinuity design approximates a randomized trial, and we introduce a novel design parameter to inform researchers about the strength of the approximation.
\end{enumerate}
\section{Methods}
\label{methods}
We define two populations of individuals: the general population (GP) and the study population (SP). We think of the GP as the global population of individuals who are susceptible to a particular disease. We think of the (SP) as a regional population of individuals who are susceptible to a particular disease. We will use subscripts to specify particular details, e.g. $\textrm{GP}_{\textrm{Crohn's}}$, $\textrm{SP}_{\textrm{Japan}}$, etc. We write CS for a clinical sample of size $n$, selected from the GP or the SP. In the former case we write $\textrm{CS}(\textrm{GP})$. In the latter case we write $\textrm{CS}(\textrm{SP})$. When the CS consists of patients presenting between some start time $t_s$ and another end time $t_e$ we write $\textrm{CS}_{[t_s,t_e]}$. For example, with times measured in years A.D., we may write $\textrm{CS}_{[2010,2020]}(\textrm{SP}_{\textrm{Japan}})$. We address inclusion, exclusion, and retention of patients in Section \ref{discussion}.

Clinicians may intervene on a sub sample of the CS. An intervention, $\tau$, is a function that takes a sub sample of individuals as an input and produces a treated sub sample of individuals as an output. The first subscript of $\tau$ specifies which treatment, and the second subscript of $\tau$ specifies how the sub sample was selected. We will analyze three, separate dietary-interventions for Crohn's disease: a plant based diet (PBD) \citep{Chiba2017}, the Mediterranean diet (MD), and the specific carbohydrate diet (SCD) \citep{Lewis2021}. We describe the design of \citet{Chiba2017} as a temporal-discontinuity (TD) design, and we describe the design of \citet{Lewis2021} as a randomized (R) design. We utilize a subscript appended to TD or R to specify the proportion $p$ of the CS that was selected for a particular intervention. For a TD design it is always the temporally latter proportion of patients within the CS that are treated. 

For each individual patient within a SP of size $N$, presentation time is modeled with a random variable $T_i$ having continuous support on $[t_s,t_e]$. We assume independence but not identical distributions, and we write $\sigma^2$ for the minimum variance over $\{T_1,...,T_N\}_{i=1}^N$. As a sample of the SP, the CS consists of the $n$ individuals $i_{j_1},...,i_{j_n}$with presentation times satisfying $t_s<t_{j_1}<...<t_{j_n}<t_e$. Define $t_{(1-p)}$ as the $(1-p)$-quantile of the distribution of $t_{j_1},...,t_{j_n}$values. The intervention $\tau_{x,\textrm{TD}_p}$ thus treats precisely those individuals in the set $\{j\in \textrm{CS}: t_{(1-p)}<t_j<t_e\}$ with treatment $x$. The resulting subset of treated individuals would be denoted with $\tau_{x,\textrm{TD}_p}(CS)$.  

To compare the effectiveness of various interventions we define a functional SR that takes as an input the GP, a SP, a CS, or a sub sample of the CS, and returns as an output the proportion of successful outcomes over the specified input population or sample. In our notation we may summarize the TD-trial of \citet{Chiba2017} with
\[\textrm{SR}(\tau_{\textrm{PBD},\textrm{TD}_{100\%}}(\textrm{CS}_{[2003,2015]}(\textrm{SP}_{\textrm{Japan}})))=96\%.\]
Likewise, we write US for the United States and summarize the randomized trial of \citet{Lewis2021} with 
\[\textrm{SR}(\tau_{\textrm{MD},\textrm{R}_{50\%}}(\textrm{CS}_{[2017,2019]}(\textrm{SP}_{\textrm{US}})))=43.5\% \quad \textrm{and} \quad \textrm{SR}(\tau_{\textrm{SCD},\textrm{R}_{50\%}}(\textrm{CS}_{[2017,2019]}(\textrm{SP}_{\textrm{US}})))=46.5\%.\]

To compare two treatments $x_j$ and $x_k$ we may--within an experiment on volunteers--assign to each individual the opposite of what they prefer. Preferences could be formed with input from medical professionals or other advisors. We assume strong and well-defined preferences, i.e. extreme propensity probabilities, and as such our methodology is complementary to the methodology described in \citet{Li2018} and \citet{Li2017}. We denote sub populations and sub samples of individuals preferring treatment $x_j$ by appending a $j$ subscript, and likewise for $x_k$ with an $k$ subscript. Assuming experimental control over SP we may deliberately assign the opposite of what is preferred to each individual and express the resulting success rates as $\textrm{SR}(\tau_{k}( \textrm{CS}_j))$ and $\textrm{SR}(\tau_{j}( \textrm{CS}_k))$. In an observational study we may observe the GP or a $\textrm{CS}(\textrm{GP})$. 
The success rate for a placebo or standard treatment observed in an observational study of a $\textrm{CS}(\textrm{GP})$ can be denoted with $\textrm{SR}(\tau_0(\textrm{CS}(\textrm{GP})))$. Success rates for preferred treatments in an observational study on $\textrm{CS}(\textrm{GP})$ are denoted with $\textrm{SR}(\tau_j(\textrm{CS}(\textrm{GP}_j)))$ and $\textrm{SR}(\tau_k(\textrm{CS}(\textrm{GP}_k)))$. 

 Even if the MD and the SCD were the only two available treatments it would be wrong to formulate an intervention policy that prescribes the SCD to everyone in the SP. All populations including the SP have some degree of heterogeneity and on this particular SP there could be patients who benefit from the MD and are harmed by the SCD. A detailed discussion of conditional interventions is provided in Appendix \ref{CI}.

Flexible intervention policy can improve success rates. We utilize deterministic counterfactual reasoning to develop a method to analyze the sensitivity of measured success rates to increasingly conditioned interventions. We consider a simple randomized trial comparing two treatments $x_j$ and $x_k$ on a large sample or population of size $n$. We write $P_k$ for the probability of the simple randomization and also for the proportion of individuals assigned to treatment $x_k$. We denote the resulting success rates with $r_j$ and $r_k$. We define the individual potential outcomes $\{Y_{ji},Y_{ki}\}_{i=1}^n$. $Y_{ji}$ is the outcome for individual $i$ should that individual select treatment $x_j$, and $Y_{ki}$ is the outcome for individual $i$ should that same individual select treatment $x_k$. Let $\alpha$ denote the proportion of the sample or population with $Y_{ji}\neq Y_{ki}$. If $x_j$ is a placebo then $\alpha$ gives the proportion of individuals affected, positively or negatively, by the treatment $x_k$. If $x_j$ and $x_k$ are separate treatments, then $\alpha$ gives the proportion of individuals who could benefit from changing from one treatment to the other. To assess the potential to increase success rates with flexible intervention policies we conduct sensitivity analyses with the following constrained optimization problems.

\begin{subequations}
\label{UL}
\begin{align}
U &= \max_{\{Y_{ji},Y_{ki}\}_{i=1}^n} \  \sum_{i=1}^n \max \{Y_{ji},Y_{ki}\}/n,\\
L &= \min_{\{Y_{ji},Y_{ki}\}_{i=1}^n} \  \sum_{i=1}^n \min \{Y_{ji},Y_{ki}\}/n, \
\textrm{both subject to}\\
\sum_{i=1}^n P_k Y_{ki}/n \ &= \ r_k \\
\sum_{i=1}^n (1-P_k) Y_{ji}/n \ &= \ r_j, \
\textrm{and optionally, when $\alpha$ is known,}\\
\sum_{i=1}^n(Y_{ji}(1-Y_{ki})+(1-Y_{ji})Y_{ki})/n&= \alpha.
\end{align}
\end{subequations}
\section{Results}
\label{results} 
\subsection{Primacy of Experimental Control Over Randomization}
\label{res1}
Let $\tau_j$ and $\tau_k$ be two separate treatments. Estimands of interest are $\textrm{SR}(\tau_j(\textrm{GP}))$ and $\textrm{SR}(\tau_k(\textrm{GP}))$. However, in an observational study of any $\textrm{CS}(\textrm{GP})$, including a representative sample or even a census, we lack experimental control and may only observe $\textrm{SR}(\tau_j(\textrm{CS}(\textrm{GP})_j))$ and $\textrm{SR}(\tau_k(\textrm{CS}(\textrm{GP})_k))$, which as estimates are susceptible to selection bias. From a randomized trial comparing $\tau_j$ and $\tau_k$, with randomization probability for $\tau_k$ equal to $p$, on a $\textrm{CS}(\textrm{SP})$, we obtain the estimates $\textrm{SR}(\tau_{j,\textrm{R}_{1-p}}(\textrm{CS}(\textrm{SP})))$ and $\textrm{SR}(\tau_{k,\textrm{R}_{p}}(\textrm{CS}(\textrm{SP})))$, which may not be externally valid on the GP. Let us assume a) that the $\textrm{CS}(\textrm{SP})$ is representative of the GP, and also b) that individuals within the $\textrm{CS}(\textrm{SP})$ are willing to honestly share their individual treatment preferences, and c) that individuals within a representative $\textrm{CS}(\textrm{GP})$ are willing to honestly share their individual treatment preferences and participate in an observational study. Write $q$ for the proportion of individuals who prefer treatment $\tau_k$. The size of the $\textrm{CS}(\textrm{SP})$ is $n$. 

The success rates of the randomized trial can be broken down conditional on treatment preference, resulting in the following four rates and respective sample sizes ($n_{jj}$, $n_{jk}$, $n_{kj}$, and $n_{kk}$): 
\begin{subequations}
\label{T}
\begin{align}
\label{T1}&\textrm{SR}(\tau_{j,\textrm{R}_{(1-p)}}(\textrm{CS}(\textrm{SP})_j)),\quad n_{jj}=(1-p)(1-q)n, \\
\label{T2}&\textrm{SR}(\tau_{j,\textrm{R}_{(1-p)}}(\textrm{CS}(\textrm{SP})_k)),\quad n_{jk}=(1-p)qn,\\ \label{T3}&\textrm{SR}(\tau_{k,\textrm{R}_p}(\textrm{CS}(\textrm{SP})_j)),\quad n_{kj}=p(1-q)n, \textrm{~and} \\
\label{T4}&\textrm{SR}(\tau_{k,\textrm{R}_p}(\textrm{CS}(\textrm{SP})_k)),\quad n_{kk}=pqn.
\end{align}
\end{subequations}

The rates in (\ref{T1}) and (\ref{T4}) could be redundant, since the larger observational study produces $\textrm{SR}(\tau_{j}(\textrm{CS}(\textrm{GP})_j))$ and $\textrm{SR}(\tau_{k}(\textrm{CS}(\textrm{GP})_k))$. Alternatively, on the same SP, the experimenter could have controlled treatment assignment not with randomization but through deliberate assignment of the opposites, resulting in the following, where we have written $n^\star$ for a smaller sample size:
\begin{subequations}
\label{TT}
\begin{align}
\label{TT1}&\textrm{SR}(\tau_j(\textrm{CS}(\textrm{GP})_j))\\
 \label{TT2}&\textrm{SR}(\tau_j(\textrm{CS}(\textrm{SP})_k)),\quad n_{jk}=qn^\star, \\
\label{TT3}&\textrm{SR}(\tau_k(\textrm{CS}(\textrm{SP})_j)),\quad n_{kj}=(1-q)n^\star, \textrm{~and}\\
\label{TT4}&\textrm{SR}(\tau_k(\textrm{CS}(\textrm{GP})_k)),
\end{align}
\end{subequations}
where we have substituted $\textrm{SR}(\tau_{j}(\textrm{CS}(\textrm{GP})_j))$ and $\textrm{SR}(\tau_{k}(\textrm{CS}(\textrm{GP})_k))$ in (\ref{TT1}) and (\ref{TT4}). We suppose that the size of $\textrm{CS}(\textrm{GP})$ is much larger than the size of $\textrm{CS}(\textrm{SP})$. Comparison of respective sample sizes across (\ref{T}) and (\ref{TT}) demonstrates that deliberate assignment of the opposites can be more efficient producing equal or more power after transport as long as the smaller sample size $n^\star$ is greater than $\max\{p,1-p\}n$. If $p=1/2$ we would need to recruit only half as many patients to participate in the experiment on $\textrm{CS}(\textrm{SP})$. Assigning the opposites is thus an efficient alternative to randomization, and our results reveal the primacy of experimental control over randomization. Further discussion of the necessity of experimental control occurs in Section \ref{discussion} and Appendix \ref{A3}.

\subsection{Formal Conditions for Equivalence Between Temporal-discontinuity and Randomized Designs}
\label{res2}
Here we apply a treatment $x$ to a proportion $p$ of a clinical sample $\textrm{CS}(\textrm{SP})$ of size $n$. We may intervene with a temporal discontinuity trial or a randomized trial on a sub sample of size $pn$. We seek conditions to ensure 
\begin{equation}\label{first}\textrm{SR}(\tau_{x,\textrm{TD}_{p}}(\textrm{CS}(\textrm{SP})))=\textrm{SR}(\tau_{x,\textrm{R}_{p}}(\textrm{CS}(\textrm{SP}))) .\end{equation}
This analysis is conditional on a fixed $\textrm{CS}(\textrm{SP})$. The randomness of the treated sub sample $\tau_{x,\textrm{TD}_{p}}(\textrm{CS}(\textrm{SP}))$ depends on uncontrolled, natural processes, sufficiently modeled with random, individual, presentation times as described in Section \ref{methods}. The randomness of the treated sub sample $\tau_{x,\textrm{R}_{p}}(\textrm{CS}(\textrm{SP}))$ depends on the controlled randomization of the randomized design. 

Let $\Omega$ denote our sub sample space. Each element $\phi\in\Omega$ is a sub sample of size $pn$. Each element $\phi$ has a probability $P(\phi)$. For the temporal discontinuity design we write $P_{\textrm{TD}_p}(\phi)$ and for the randomized design we write $P_{\textrm{R}_p}(\phi)$. Associated with $P_{\textrm{R}_p}(\phi)$ is a uniform probability distribution $\mu_{\textrm{R}_p}$ over $\Omega$. Associated with $P_{\textrm{TD}_p}(\phi)$ is an unknown probability distribution $\mu_{\textrm{TP}_p}$ over $\Omega$. When the two distributions are equal, i.e. when  
\begin{equation}\label{second}\mu_{\textrm{R}_p}=\mu_{\textrm{TD}_p},\end{equation}
we say that the designs are equivalent. We refer to the equality in (\ref{second}) as our \emph{equivalence condition}. Note that (\ref{second}) implies (\ref{first}).

As described in Section \ref{methods}, a carefully designed temporal-discontinuity trial specifies a start time $t_s$, a cut time $t_{(1-p)}$, and an end time $t_e$. The individuals who present between $t_{(1-p)}$ and $t_e$ are treated. For each individual, presentation time is a random variable, and the minimum of the individual variances of those presentation times, computed over the whole SP, is denoted with $\sigma^2$. Define the following design parameter: \begin{equation}\label{K}K:=\sigma^2/(t_e-t_s).\end{equation}
When $K$ is large then (\ref{second}) is approximately satisfied, and the temporal-discontinuity design approximates a randomized design. 
\subsection{Assessing the Sensitivity of Success Rates to Departures From the Assumption of Homogeneous Effects}
\label{res3}
The constrained optimization problems in (\ref{UL}) have solutions that depend not on the simple randomization probability $P_k$ but only on the success rates $(r_j,r_k)$ and the optional parameter $\alpha$, which is the proportion of individuals who could benefit from changing treatments (see Section \ref{methods}). When $\alpha$ is known the solutions of the problems in (\ref{UL}) are given by \begin{equation}\label{sol1}U(\alpha)=\min\left\{\frac{(r_j+r_k)+\alpha}{2},1\right\} \textrm{~and}\end{equation}
\begin{equation}\label{sol2}
L(\alpha)=1-\min\left\{\frac{(1-r_j)+(1-r_k)+\alpha}{2},1\right\}.\end{equation}
Given $r_j$ and $r_k$ the domain of $U(\alpha)$ is the interval $I_U=[|r_j-r_k|,\min\{r_j+r_k,1\}]$. We see from (\ref{sol1}) that the maximum possible success rate $U(\alpha)$ ranges linearly from a low of $\max\{r_j,r_k\}$ up to a high of $\min\{r_j+r_k,1\}$. Likewise, given $r_j$ and $r_k$ the domain of $L(\alpha)$ is the interval $[|(1-r_j)-(1-r_k)|,\min\{(1-r_j)+(1-r_k),1\}]$. We see from (\ref{sol2}) that the minimum possible success rate $L(\alpha)$ ranges linearly from a high of $1-\max\{1-r_j,1-r_k\}=\min\{r_j,r_k\}$ down to a low of $1-\min\{(1-r_j)+(1-r_k),1\}$. When $\alpha$ is unknown success rates could be as low as \begin{equation}\label{sol3}1-\min\{(1-r_j)+(1-r_k),1\}\end{equation} and as high as \begin{equation}\label{sol4}\min\{r_j+r_k,1\}.\end{equation} Further details and mathematical proofs are provided in Appendix \ref{A1}.  
\section{Application}
\label{application}
There are fundamental reasons to believe that diet plays an important role in Crohn's disease. Surgery to exclude the colon from intestinal transit resolves inflammation, but the restoration of transit results in recurrence of disease \citep{Rut}. It is not clear whether the inflammation arises from a constituent of the bowel contents or an abnormal immune reaction to normal bowel contents \citep{Mark}. However, there are exclusion diets and other diets that are moderately effective as therapies for Crohn's disease \citep{Wicek2022}. Interestingly, transfer of gut microbiota in mice can facilitate colitis \citep{Gao2018}.

Throughout this example application we will consider three dietary therapies as treatments for active Crohn's disease. The considered diets are the specific carbohydrate diet (SCD), the Mediterranean diet (MD), and a plant based diet (PBD). We are concerned primarily with formulating an intervention policy that is applicable to a general population (GP) of patients with Crohn's disease. The primary outcome is remission. The definition of remission is discussed in Section \ref{discussion}. We focus mainly on evidence from the randomized trail of \citet{Lewis2021} and the quasi temporal-discontinuity trial of \citet{Chiba2017}. 

\citet{Lewis2021} has compared the effectiveness of MD vs SCD by utilizing randomized treatment assignment. Their study population (SP) of patients was located within the United States (US), and we denote it with $\textrm{SP}_{\textrm{US}}$. \citet{Chiba2017} has had high levels of success with a plant-based dietary intervention, but their report summarizes a ``single-group, nonrandomized, open noncontrolled trial.'' Their study population of patients was located in Japan, and we denote it with $\textrm{SP}_\textrm{Japan}$. The interventions are $\tau_{\textrm{MD},\textrm{R}_{50\%}}$, $\tau_{\textrm{SCD},\textrm{R}_{50\%}}$, and $\tau_{\textrm{PBD},\textrm{TD}_{50\%}}$. The first two interventions are applied to a $\textrm{CS}_{[2017-2019]}(\textrm{SP}_{\textrm{US}})$ and the third intervention is applied to a $\textrm{CS}_{[2003,2015]}(\textrm{SP}_{\textrm{Japan}})$. The success rates (SRs) can be summarized as follows:
\begin{align}\label{SRs1}&\textrm{SR}(\tau_{\textrm{MD},\textrm{R}_{50\%}}(\textrm{CS}_{[2017,2019]}(\textrm{SP}_{\textrm{US}})))=43.5\%, \quad \textrm{SR}(\tau_{\textrm{SCD},\textrm{R}_{50\%}}(\textrm{CS}_{[2017,2019]}(\textrm{SP}_{\textrm{US}})))=46.5\%, \textrm{~and} \\ \label{SRs2} & \textrm{SR}(\tau_{\textrm{PBD},\textrm{TD}_{100\%}}(\textrm{CS}_{[2003,2015]}(\textrm{SP}_{\textrm{Japan}})))=96\%.\end{align}

For simplicity we have ignored sampling error and the stratified randomization of \citet{Lewis2021}. If we consider only the success rates from the randomized trial it appears that the SCD intervention is optimal with a success rate of $46.5\%$. However, Crohn's disease presentation and progression is highly heterogeneous \citep{Furey2019}. Some patients may do better with the SCD than with the MD, while other patients may do better with the MD than with the SCD. Applying the result of Section \ref{res3} as described in (\ref{sol3}) to the success rates $r_\textrm{MD}=43.5\%$ and $r_\textrm{SCD}=46.5$ demonstrates the potential of flexible interventions to increase the success rate up to $90\%$. Conditional interventions are defined and discussed in Appendix \ref{CI}. That potential is positively and linearly related to $\alpha$ (see Section \ref{methods} and (\ref{sol1})). That potential would be less if it were demonstrable that a large proportion of individuals were destined for remission or non remission regardless of intervention (SCD or MD).

The success rate reported by \citet{Chiba2017} is still higher at $r_\textrm{PBD}=96\%$. As we just saw, one way to increase a success rate is to conditionally intervene on patients who will probably benefit from the treatment; see (\ref{UL}), (\ref{sol1}) and (\ref{sol3}). It is possible that the high success rate reported by \citet{Chiba2017} is due to selection bias \citep{Lu2022,Smith2019}. Restated in our notation, it is possible in (\ref{SRs2}) that $\textrm{CS}_{[2003,2015]}(\textrm{SP}_{\textrm{Japan}})$ is not representative of $\textrm{SP}_{\textrm{Japan}}$. Remarkably, there is no guarantee in (\ref{SRs1}) that $\textrm{CS}_{[2017,2019]}(\textrm{SP}_{\textrm{US}})$ is representative of $\textrm{SP}_{\textrm{US}}$, and in neither (\ref{SRs1}) nor (\ref{SRs2}) do we have a guarantee that the SPs are representative of the GP. However, we should compare the unbiased $\tau_{\cdot,\textrm{R}_{50\%}}$ operator of (\ref{SRs1}) with the possibly biased $\tau_{\textrm{PBD},\textrm{TD}_{50\%}}$ operator of (\ref{SRs2}).

We have written $\textrm{TD}_{100\%}$ in (\ref{SRs2}) because the study by \citet{Chiba2017} is missing a control group. Presumably, there was a time before their clinic was intervening with a PBD, but more than a decade ago \citet{Chiba2010} reported successful intervention with a PBD, and as we have been mentioning their time frame is $[2003,2015]$. Perhaps other Japanese trials could serve as a control. There are Japanese studies without a plant-based dietary intervention that have lower levels of success \citep[Table 4]{Chiba2017}. See \citet{Rosenbaum2010} for further reading on evidence factors in observational studies. Another technique with utility in this context is the difference in differences technique described by \citet{Wing2018}.

Ideally, a temporal-discontinuity trial would treat the first half of a sequence of patients with a placebo or conventional treatment and the second half of the same sequence of patients with an improved treatment, in which case we would write $\textrm{TD}_{50\%}$. Also it would be ideal if the time frame of the study was compressed around the cut point; see Appendix \ref{A3} where we discuss discontinuity designs. In particular, the noise induced randomization design of \citet{eck} appears applicable, especially since ``patients with Crohn's disease often have symptoms for several years before the correct diagnosis is confirmed" \citep{Cushing2021}. In this scenario a large $\sigma^2$ value (see Section \ref{methods}) is thus plausible, which is supportive of causal inference, but the length of time of the study by \citet{Chiba2017}, $t_e-t_s=12$ years, is too long, so that the design parameter $K$ of Section \ref{res2} is not large enough to support an argument for approximation of a randomized design.

However, the PBD was developed in response to perceived Westernization of dietary patterns in Japan \citep{Chiba2019}. Therefore, from a whole-of-Japan perspective, the PBD appears similar to assigning the opposites; see Sections \ref{methods} and \ref{res1}. A subset of the patients described in \citet{Chiba2017} were ``consecutive adults", and it would have been possible to ask those adults for their honest dietary preferences. Data from a sub sample of patients who prefer a Westernized diet but instead consented to the PBD would therefore be efficiently transported to wider populations as described in Section \ref{results}. In Section \ref{discussion} we provide further discussion to improve the design of studies like the study of \citet{Chiba2017}.
\section{Discussion}
\label{discussion}
In this paper we have shown how assigning to each individual the opposite treatment of that which is preferred by each individual is efficient for transport. We have also described some conditions that facilitate assessment of whether a temporal-discontinuity trial well approximates a randomized trial. And we have developed a method of sensitivity analysis to assess the potential of flexible intervention policy to increase success rates. We have applied those techniques to analyze which of three dietary interventions are to be recommended to supplement conventional treatment for Crohn's disease. We are not in a position to make a definitive statement about which dietary intervention is best for most individuals. However, we have demonstrated in a variety of ways that it would be less than optimal to simply discard all evidence that does not arise from randomized experiments, and it would be less than optimal to formulate a universal prescription for all individuals within the general population to accept the treatment with the highest success rate from some well designed randomized trial.

Doing the opposite may improve efficiency of transport. Randomized trials of treatments thought to affect rare outcomes may be under powered. For instance, an early randomized controlled trial of the BNT162b2 mRNA Covid-19 Vaccine randomly assigned $18,860$ individuals to treatment and $18,846$ to placebo \citep{Polack2020}. Months of follow-up demonstrated effectiveness of the vaccine to prevent cases of Covid-19 and also severe cases of Covid-19. However, no Covid-19-associated deaths were observed in either the treatment or placebo groups. The number of participants was too small. 

Efficiency gains may be limited to the context of open-label trials where participants have knowledge of the assigned treatment. A placebo-controlled trial that assigns the opposites may be possible, but there are ethical considerations \citep{Herrera2001}. Deception may be required to control for the placebo effect and assign the opposite. Other limitations include the requirements of well-defined preferences and honesty of the participants. Also, the redundancy of randomization can be a strength in that the data from the redundant groups can be compared with relevant observational data to check the assumption of external validity.

Here we briefly discuss exclusion or attrition of patients. If any of the results of \citet{Chiba2023} were obtained from a sample that excluded $u\%$ of $\textrm{CS}(\textrm{SP}_{\textrm{Japan}})$ and then $v\%$ of the treated group withdrew from treatment protocol then adjustment would lower the success rate by at most $(u+v)\%$. A subset of the patients described in \citet{Chiba2017} were ``consecutive adults", and two of those adults withdrew due to intestinal obstruction, resulting in a still impressive remission rate of $24/26\approx 92\%$ amongst those adults \citep{Chiba2017}[Table 2]. Also, while it is true that different studies define remission differently, it is also true that the plant-based dietary intervention had higher rates of success for secondary outcomes including long term remission and avoidance of surgery \citep{Chiba2023}[Figure 14, Table 5].

Randomization balances observed and unobserved covariates \citep{Hariton2018}, but idealized natural experiments can do the same \citep{eck}. There may be sufficient randomness for causal inference from observational data, but in the absence of control by an experimenter there may be concerns about the stable unit treatment value assumption \citep[Section 5.2]{Knaeble2023}. This subtle issue is discussed further in Appendix \ref{A3}, where additional references are provided. We state the main point here within the context of the PBD of \citet{Chiba2023}. Even if the PBD is randomly assigned there is a distinction between an experimenter knowingly assigning just the PBD and a natural process stochastically assigning not just the PBD but also possibly other more subtle treatments like exercise or psychological support from a devoted nurse. The clinic that naturally prescribes a PBD could also be the clinic that intervenes in other healthy ways, unless we have assurance of experimental control in the form of an experimenter who testifies that they deliberately assigned only the PBD as part of an experiment.

If multiple clinics conduct TD trials of the same causal relationship but with differing temporal cut values, and each reports similar estimates of the causal effect, then that would support a technical assumption about continuity of treatment effects, as described in Section \ref{A3}, and warrant policy formulation on a better defined population. If the cutoff values were independently and randomly selected across studies then that would further facilitate policy formation. For further reading on multiple cutoff values see \citet{ZhangSafe}. 

Since patients will most often present at a clinic nearby their residence, a merged temporal-discontinuity design may be utilized. The merged TD design may include a variety of geographically dispersed clinics, each implementing a temporal cut point of their choosing. But, within the merged SP, while we would include patients from many clinics, at each clinic we would include only those patients presenting relatively close in time to the chosen cut point of the clinic at which they present. This would better approximate a randomized trial; c.f. the design parameter $K$ of Section \ref{res2}.

In summary, it is not necessarily a lack of randomization alone that dissuades belief in causal inference. It is primarily a lack of control or lack of knowledge regarding how that control was implemented to assign treatments. We have discussed randomized trials and temporal-discontinuity trials, and demonstrated the utility of both to support evidence-based formulation of intervention policy. We view the randomized design and the temporal-discontinuity design as complementary designs. If a conditional intervention is being applied to a sub population of patients each with a ``clinical consensus that is ambiguous or divided'' \citep{Li2017} then supporting evidence may be found in transportable results from other randomized trials. However, if the conditional intervention is conceived as an act of interference on a sub population of patients with extreme treatment propensity probabilities, then intervention policy may be best informed by the more efficiently transportable results of controlled experiments that deliberately assign the opposite treatment to that desired by participants.
\section*{Acknowledgement}
Thank you to Samuel Whitlock for conceiving and sharing the idea to efficiently assign treatments opposite to preferences. 
\clearpage
\appendix
\section{Mathematical Proofs}
\label{A1}
Define  
\begin{subequations}
\label{counts}
\begin{align}
&s=\sum_{i=1}^n Y_{ki}Y_{ji}/n, \\
&t=\sum_{i=1}^n (1-Y_{ki})(1-Y_{ji})/n,\\ &u=\sum_{i=1}^n Y_{ki}(1-Y_{ji})/n, \textrm{~and} \\
&v=\sum_{i=1}^n (1-Y_{ki})Y_{ji}/n,
\end{align}
\end{subequations}
and restate the optimization problem in (\ref{UL}), without the optional constraint, and only the $U$ objective function, as
\begin{subequations}
\label{AppSub}
\begin{align}
\label{A}
U = \max_{\{s,t,u,v\}} (s+u+v)&= 1-\min_{\{s,t,u,v\}} t , \
\textrm{subject to}\\
\label{B}0&\leq s,t,u,v\leq 1,\\
\label{C}s+t+u+v&=1, \\
\label{D}s+u &=r_k, \textrm{~and} \\
\label{E}s+v &=r_j.
\end{align}
\end{subequations}
By adding (\ref{D}) and (\ref{E}) we obtain \begin{equation}\label{F}2s+u+v=r_j+r_k.\end{equation} Subtracting (\ref{F}) from (\ref{C}) results in \begin{equation}\label{G}t-s=1-(r_j+r_k).\end{equation} Reasoning from (\ref{G}) and (\ref{B}), to minimize $t$ we push $s$ to its minimum and set \[s=\max\{(r_j+r_k)-1,0\},\] resulting in 
\[t=1+\max\{r_j+r_k-1,0\}-(r_j+r_k).\] From there via (\ref{A}) we obtain
\begin{align*}
U=1-t&=1-(1+\max\{r_j+r_k-1,0\}-(r_j+r_k))\\
&=\min\{1-(r_j+r_k),0\}+(r_j+r_k)\\
&=\min\{r_j+r_k,1\}.
\end{align*}
By symmetry we obtain
\[L=1-\min\{(1-r_j)+(1-r_k),1\}.\] 

Denote the proportion of the population that is affected by the treatment with 
\begin{equation}\label{W}\alpha:=u+v.\end{equation}
From (\ref{E}) and (\ref{D}) we have 
\begin{equation}\label{X}|r_j-r_k|\leq \alpha \leq \min\{r_j+r_k,1\}.\end{equation}
From (\ref{C}) and (\ref{W}) we obtain
\begin{equation}\label{Y}t+s=1-\alpha.\end{equation}
If $\alpha$ is known then the equation in (\ref{Y}) is an optional constraint for the problem described in (\ref{AppSub}).
Adding (\ref{G}) and (\ref{Y}) results in
\[t=\max\left\{\frac{2-(r_j+r_k)-\alpha}{2},0\right\}\]
from which we obtain
\[U(\alpha)=\min\left\{\frac{(r_j+r_k)+\alpha}{2},1\right\}.\]
Likewise, 
\[L(\alpha)=1-\min\left\{\frac{(1-r_j)+(1-r_k)+\alpha}{2},1\right\}.\]
Given $(r_j,r_k)$ the function $U(\alpha)$ linearly maps its domain $[|r_j-r_k|,\min\{r_j+r_k,1\}]$ (see (\ref{X})) onto a space of success rates ranging from $\max\{r_j,r_k\}$ up to $\min\{r_j+r_k,(r_j+r_k+1)/2,1\}=\min\{r_j+r_k,1\}$, and the function $L(\alpha)$ linearly maps its similar domain $[|(1-r_j)-(1-r_k)|,\min\{(1-r_j)+(1-r_k),1\}]$ onto a space of success rates ranging from $1-\max\{1-r_j,1-r_k\}$ down to $1-\min\{(1-r_j)+(1-r_k),((1-r_j)+(1-r_k)+1)/2,1\}=1-\min\{(1-r_j)+(1-r_k),1\}$.
\section{Discontinuity Designs}
\label{A3}
Causal inference from observational data is a challenging problem. Natural experiments are one way to approach the problem. In particular, the methodology of \citet{eck} is ideal, assuming an infinite population. They describe noise-induced randomization with a discontinuity design. There is a measured running variable $Z=U+\epsilon$, where $U$ is the real latent variable and $\epsilon$ is measurement noise. Treatment is assigned to those individuals with $Z\geq c$ for some cutoff value $c$ and withheld from those individuals with $Z<c$. For small $\delta>0$ we may restrict our attention to the sub population of individuals with $Z\in(c-\delta,c+\delta)$. In the limit as $\delta\to 0$ treatment assignment is haphazard and in most circumstances it is therefore reasonable to conclude that all covariates are balanced and that causal inference is warranted.

An example will provide context and clarity. We adapt the methodology and notation of \citet{eck} and apply it to the similar study of low birth weight infants conducted by \citet{Almond10}. The threshold of very low birth weight is $c=1,500$ grams. Infants born below the threshold are treated with intensive care, while infants born above the threshold are treated with standard care. It has been observed that infants born with birth weights just above $1,500$ grams have an infant mortality rate of approximately $5.5\%$, while infants born with birth weights just below $1,500$ grams have an infant mortality rate of approximately $4.5\%$ \citep{Almond10}. Ideally, we have $Z=U+\epsilon$ where $Z$ is the measured birth weight, $U$ is the actual birth weight, and $\epsilon$ is exogenous measurement error. The latent variable $U$ is a potential confounder. However, provided with a large enough population, on the sub population of infants with $Z\in(c-\delta,c+\delta)$, in the limit as $\delta\to 0$, it may be reasonable to conclude that each strata of $U$ will be equally balanced above and below $c$. With exogenous noise $\epsilon$ and treatment actively assigned from $Z<c$ where $Z=U+\epsilon$, we conclude that balance of $U$ implies balance of all unmeasured covariates thus warranting causal inference from this noise-induced natural experiment. 

It is worth discussing a subtle point relating to that warrant for causal inference from a natural experiment. If a single treatment were not actively assigned by an experimenter with control over the situation then we would not be able to conclude that all unmeasured covariates were balanced. Suppose for instance in the context of a natural, natural experiment \citep{Rosen} nature assigns multiple exposures. Consider as the treatment the primary exposure, but note that there is no way to know whether nature balanced the secondary exposures or not. This subtle point is discussed in further detail in \citet[Section 5.2]{Knaeble2023}. It is related to the stable unit treatment value assumption \citep[p. 10]{Imbens2015} and also the exclusion restriction \citep[Section 3.2]{Angrist1996}. Similar concerns about compound treatments arise in randomized experiments \citep{Hernan}.

The noise-induced discontinuity design is a clever approach, but its results are not as compelling as the results of an actual experiment conducted on a representative sample of a well defined population. The problem becomes apparent when it is realized that the noise-induced discontinuity design requires conditioning on the results of a noisy process. Recall how $Z=U+\epsilon$. Suppose continuous probability density functions $h$ and $g$ for $U$ and $\epsilon$, respectively. The probability density function $h$ represents the population distribution of $U$ values. The probability function $g$ represents exogenous measurement error, assumed to be independent and identically distributed across individuals. The sub population of individuals with $Z=c$ have a $U$ distribution with a density function proportional to $h(u)g(u-c)$. If we assume relatively small variance of the noise then $h(u)$ is roughly constant near $c$ and conditional on $Z=c$ we have a distribution of $U$ values with probability density function approximately equal to $g(u-c)$. For instance, if the measurement noise is Gaussian then we have a warrant for causal inference on a sub population of individuals with $U$ values normally distributed about the cutoff $c$. But that is not a representative sample of any sub population well defined from $U$. If we allow heterogeneous causal effects that depend on $U$ then our causal inference could be biased unless the noise term is uniformly distributed.

\begin{figure}[h!]
\centering 
\includegraphics[width=3.8in]{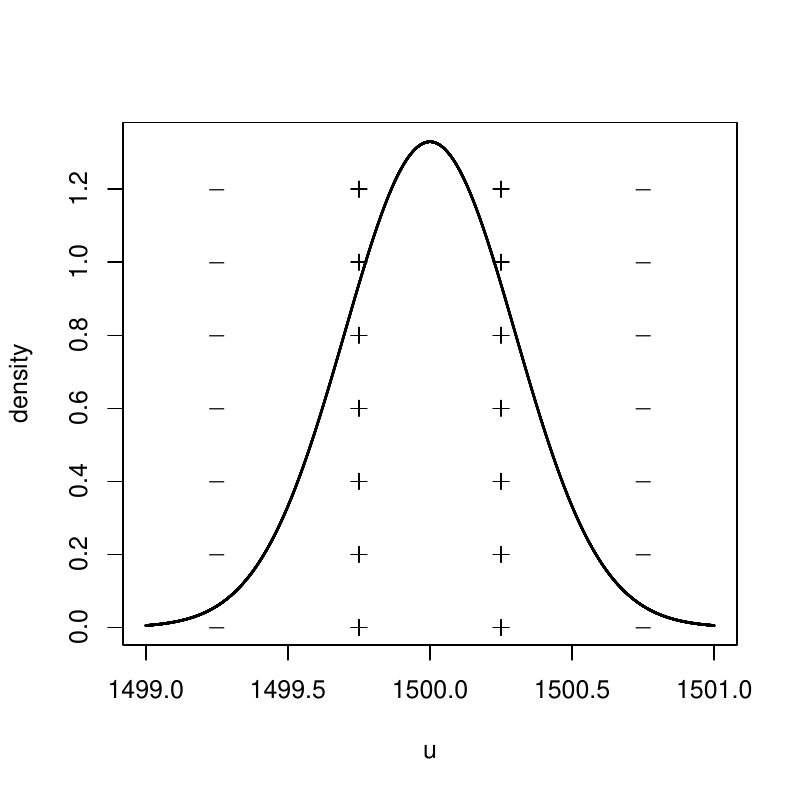}
\caption{With Gaussian noise we observe a Gaussian distribution of $U$ values conditional on $Z=c=1,500$ which could result in an over selection of individuals who benefit from treatment (indicated with plus symbols in the plot) and an under selection of individuals who are harmed by treatment (indicated with minus symbols in the plot), warranting causal inference on the sub population with $1,499<u<1,501$ while the average treatment effect on that same sub population is zero, limiting the effectiveness of any policy recommendation for the treatment across the whole sub population.} 
\label{BellPlot}
\end{figure}

This is a subtle point worth restating in the language of our example involving low birth weight infants. Yes, we can conduct causal inference; we can conclude that the treatment of intensive care provided to very low birth weight infants saved lives. No, we can not establish an intervention policy on a well defined sub population and estimate the average effect of our intervention, if the noise distribution is anything but uniform. We can't say for example that intensive care saves approximately $1\%$ of infants with actual birth weights between $1,499$ and $1,501$ grams, because there are alternate explanations on that particular sub population. It could be for example with Gaussian noise that the treatment of intensive care saved lives on the over sampled segment right near the cutoff of $c=1,500$ but lost lives on the under sampled segments slightly further away from the cutoff near $c=1,499$ grams and $c=1,501$ grams. That scenario could be consistent with both a) the observed data that arose from a non-representative sample conditional on $Z$ which contains noise and b) a smaller or non-existent average treatment effect on that sub population with $1,499<u<1,501$, depending on the variance of the noise; see Figure \ref{BellPlot}. While a continuity or monotonicity of effects argument could be made, the thought experiment just described in this section reveals the importance of a well-defined population for intervention policy.
\section{Conditional Interventions}
\label{CI}
\citet{Pearl2009} represents interventions with the ``do operator''. Within that framework an intervention is a ``hypothetical situation in which treatment ... is administered uniformly to the population'' \citep{Pearl2009}. That definition of an intervention means that all individuals of a population have their treatment variable value set to the same prescribed value. Those individuals within the population whose treatment variable status was already at that prescribed value remain unaffected by such an intervention.

However, commenting on the use of causal diagrams for empirical research, \citet{ImbRubDis} have written the following. \begin{quote} ``Important subject-matter information is not conveniently represented by conditional independence in models with independent and identically distributed random variables. Suppose that, when a person's health status is 'good', there is no effect of a treatment on a final health outcome, but, when a person's health status is 'sick', there is an effect of this treatment, so there is dependence of final health status on treatment received conditional on initial health status. Although the available information is clearly relevant for the analysis, its incorporation, although immediate using potential outcomes, is not straightforward using graphical models.''\end{quote} There are graphical models that unify causal graphs and potential outcomes \citep{SWIGs,Malinsky2019-gd,Pearl2010-ue,Dawid2002-uz}, but here we will utilize potential outcomes notation.

Merriam-Webster's dictionary defines a medical intervention as an ``act of interfering with the outcome or course especially of a condition or process (as to prevent harm or improve functioning)'' \citep{MW}. We emphasize that an intervention is an act of interfering, and we proceed to define conditional interventions as follows, so as to preclude consideration of the contradictory idea of intervening on individuals who do what they already planned to do. 

Consider those individuals who plan on treatment $X=x$. An intervention conditional on $x$ is the act of interfering and assigning $X=(x+k)$, for some $k\neq 0$, to each individual who plans on treatment $X=x$. This is the same as Pearl's definition previously described but applied only to individuals with $X=x$. However, we conceptualize this conditional intervention as doing the change $k$ rather than doing the changed treatment value $x+k$. Conditional interventions can be applied to sub populations who would benefit, and the notation we just introduced helps to describe a planned intervention on individuals at various and specific times. When an individual, e.g. a person, is repeatedly measured at different points in time, it is recommended to consider the measurements as arising from different experimental units \citep{Imbens2015}. The simplest case of a conditional intervention arises in a cross sectional study with a binary treatment indicator $X=0$ for control and $X=1$ for treatment. In that simplest case we may apply an intervention of $k=1$ to those who prefer control and an intervention of $k=-1$ to those who prefer the treatment. 

Suppose that $X$ causes $Y$ and we have observed the data of Table \ref{TableA}. The causal definition of no-confounding is 
$P(y|do(x))=P(y|x)$ \citep{PearlBook}. In potential outcomes notation, for $x=2$, the causal definition of no-confounding asserts that $E(Y_{x=2})=E(Y|x=2)$, which is $1/2$ in Table \ref{TableA}. The causal definition of no-confounding is satisfied. However, we could have $E(Y_{x=2}|x=0)=0$ and $E(Y_{x=2}|x=1)=1$, so the causal definition of no-confounding is technically still satisfied, but there is clearly causality apparent in the conditional, potential outcomes, and no association in the observed data. This simple example is related to what has been described as unfaithfulness of a population to a causal graph \citep[Section 4]{Sch}. To avoid the assumption of faithfulness it is therefore reasonable to introduce a causal definition of conditional no-confounding to pair with our definition of conditional interventions. Basically, conditional no-confounding means $E(Y_{x+k}|x)=E(Y|x+k)$. We can condition on more than just the treatment variable $X$, as theoretically considered implicitly in the formulation of the constrained optimization problem in (\ref{UL}).
\begin{table}[ht!]
\centering
\caption{Hypothetical, observed data on categorical variables $X$ and $Y$ showing relative frequencies and no association between $X$ and $Y$.}
\label{TableA}
\begin{tabular}{rccc}
&X=0&X=1&X=2\\
\hline
Y=1&1/6&1/6&1/6\\
Y=0&1/6&1/6&1/6\\
\hline
\end{tabular}
\end{table}
\bibliographystyle{plain}
\bibliography{refs.bib}

\end{document}